\documentclass{aa}  

\usepackage{graphicx}
\usepackage{txfonts}
\usepackage{url}
\usepackage{natbib}
\bibpunct{(}{)}{;}{a}{}{,}
\usepackage{enumitem}
\usepackage{array}
\usepackage{comment}
\usepackage[dvipsnames]{xcolor}  
\usepackage{ulem}    

\newcommand{\rs}{RS\,Oph} 

\newsavebox{\myimage}

\begin{document}

   \title{High-resolution imaging of the evolving bipolar outflows in symbiotic novae: The case of the RS\,Ophiuchi 2021 nova outburst}


  \author{R.~Lico\inst{1,2}\fnmsep\thanks{Email: rlico@ira.inaf.it}, M.~Giroletti\inst{1}, U.~Munari\inst{3}, T.~J.~O'Brien\inst{4}, B.~Marcote\inst{5,6}, D.~R.~A.~Williams\inst{4}, J.~Yang\inst{7}, P.~Veres\inst{8}, and P.~Woudt\inst{9}.} 

   \institute{INAF Istituto di Radioastronomia, via Gobetti 101, 40129 Bologna, Italy.
   \and Instituto de Astrof\'{\i}sica de Andaluc\'{\i}a-CSIC, Glorieta de la Astronom\'{\i}a s/n, 18008 Granada, Spain.
   \and INAF Osservatorio Astronomico di Padova, 36012 Asiago, VI, Italy.
   \and Jodrell Bank Centre for Astrophysics, School of Physics and Astronomy, University of Manchester, Manchester M13 9PL,UK.
   \and Joint Institute for VLBI ERIC, Oude Hoogeveensedijk 4, 7991~PD Dwingeloo, The Netherlands.
   \and ASTRON, Oude Hoogeveensedijk 4, 7991~PD Dwingeloo, The Netherlands.
   \and Dept.~of Space, Earth and Environment, Chalmers University of Technology, Onsala Space Observatory, 43992 Onsala, Sweden.
   \and Center for Space Plasma and Aeronomic Research (CSPAR), University of Alabama in Huntsville, Huntsville, AL 35899, USA.
   \and Department of Astronomy University of Cape Town, Private Bag X3, Rondebosch 7701, South Africa.
              }

   \date{Received July 3, 2024; accepted November 2, 2024}

 
  \abstract
   {The recurrent and symbiotic nova RS Ophiuchi (\rs{}) underwent a new outburst phase during August 2021, about 15 years after the last event that occurred in 2006. This outburst represents the first nova event ever detected at very high energies (VHE, E>100\,GeV), and a whole set of coordinated multiwavelength observations were triggered by this event.}
   {The main goals of this work are to characterize the evolving morphology of the expanding bipolar ejecta with high accuracy and to determine the physical conditions of the surrounding medium in which they propagate.}
   {By means of high-resolution very long baseline interferometry (VLBI) radio observations, we monitored \rs{} with the European VLBI Network (EVN) and e-MERLIN at 1.6 and 5\,GHz during multiple epochs from 14 to 65 days after the explosion.}
   {We reveal an evolving source structure consisting of a central and compact core and two elongated bipolar outflows expanding on opposite sides of the core in the east-west direction. The ejecta angular separation with time is consistent with a linear expansion with an average projected speed of $\sim7000$ km s$^{-1}$. We find clear evidence of a radial dependence of the density along the density enhancement on the orbital plane (DEOP), going from 1$\times$10$^7$ ~cm$^{-3}$ close to the central binary to 9$\times$10$^5$~cm$^{-3}$ at $\sim400$~AU.}
   {Thanks to the accurate source astrometric position provided by \textit{Gaia} DR3, in this work we draw a detailed scenario of the geometry and physics of the \rs{} evolving source structure after the most recent nova event. We conclude that most of the mass lost by the red giant companion goes into the DEOP, for which we estimate a total mass of $6.4 \times 10^{-6} ~~\mathrm{M_\odot}$, and into the circumstellar region, while only a small fraction (about one-tenth) is accreted by the white dwarf.}


\authorrunning{R. Lico et al.}
\titlerunning{Tracking the expansion of RS Ophiuchi bipolar ejecta during the 2021 nova outburst.}

   \maketitle
   
%

\section{Introduction} 
RS Ophiuchi (hereafter \rs) is a well-known recurrent and symbiotic nova located at 2.68$\pm$0.17 kpc distance according to \textit{Gaia} DR3 parallax \citep{Gaia2016, Gaia2023}.  It consists of a massive white dwarf (WD) and a M0III red giant (RG) in a close binary system with an orbital period of $\sim$$454$ days \citep{Brandi2009}.  Before accumulating on its surface, the material accreted from the RG by the WD flows through a massive accretion disk.  The near-Chandrasekhar mass of the WD coupled with its large accretion rate fit with theoretical expectations of a nova outburst every $\sim$20 yrs \citep[cf.][]{Nelson2011}, and at least seven eruptions have been already recorded from \rs, in 1898, 1933, 1958, 1967, 1985, 2006, and 2021. All of these events evolved in a nearly identical manner at optical wavelengths. During an outburst of \rs, most of the accreted material is ejected \citep[$\sim$$10^{-6} M_{\odot}$;][]{Das2006,Eyres2009,Pandey2022} at high velocity \citep[$V_{\rm ej}$$\geq$7500 km s$^{-1}$;][]{Munari2022}, and the amount of accreted mass retained on the surface of the WD is sufficient to support stable nuclear burning for about three months \citep{Osborne2011,Page2022,Munari2022a}. The accretion disk is destroyed by the nova ejecta, and it takes $\sim$260 days for it to reform and return to pre-outburst brightness \citep{Zamanov2022, Munari2022b}, marking the start of a new accretion cycle that leads to the next outburst.

\begin{table*}[!h]
\begin{center}
\begin{tiny}
\caption{Details of the observations.}
\label{table_obs}   
\setlength{\tabcolsep}{3.2pt}
\renewcommand{\arraystretch}{1.0}     
\begin{tabular}{clcccccc}  
\hline
\noalign{\smallskip}
Epoch & Obs. date & $T - T_0$ & Freq.  & Restoring beam & Map peak & $1\sigma$ rms & Stations\tablefootmark{a} \\
      & (2021) & (days) & (GHz) & (mas $\times$ mas, $^{\circ}$) & mJy/beam & mJy/beam\\
\noalign{\smallskip}
\hline    
\noalign{\smallskip}
I   & Aug. 22  & 14 & 5   & $1.5 \times 4.2 $, 75.9  & 15.66 & 0.08 & JB1,WB,EF,MC,IR,YS,TR,HH,SH \\
    & Aug. 23  & 15 & 1.6 & $4.4  \times 18.1$, 76.3 & 45.37 & 0.19 & JB1,WB,EF,MC,IR,TR,HH \\
II  & Sep. 1   & 24 & 5   & $6.9 \times 11.5$, -5.7  & 9.58  & 0.01 & JB1,WB,EF,MC,IR,YS,O8,T6 \\
    & Sep. 2   & 25 & 1.6 & $10.4 \times 43.4$, 10.7 & 18.32 & 0.07 & JB1,WB,EF,MC,IR,TR,HH,O8,T6 \\
III & Sep. 11  & 34 & 5   & $1.3 \times 2.8 $, 75.4  & 3.25  & 0.03 & JB2,WB,EF,MC,IR,YS,TR,HH,SH,O8,SV,ZC,BD,e-MERLIN \\
    & Sep. 12  & 35 & 1.6 & $9.0  \times 11.9$, 26.3 & 6.85  & 0.06 & JB2,WB,EF,MC,IR,TR,HH,O8,T6,SV,ZC,BD,e-MERLIN \\
IV  & Sep. 26  & 49 & 5   & --- & --- & --- & ---  \\
    & Sep. 27  & 50 & 1.6 & --- & --- & --- & ---  \\
V   & Oct. 11  & 64 & 5   & $1.3 \times 2.6 $, 71.3 & 4.50  & 0.04 & JB2,WB,EF,MC,IR,YS,TR,HH,SH,O8,SV,ZC,BD,NT,e-MERLIN \\
    & Oct. 12  & 65 & 1.6 & $9.9 \times 45.6$, -1.3 & 4.56  & 0.05 & JB2,WB,EF,MC,IR,TR,HH,O8,NT,SV,ZC,BD,e-MERLIN \\\noalign{\smallskip}
\hline
\end{tabular}
\end{tiny}
\tablefoot{
\begin{tiny}
\newline
\tablefoottext{a}{EVN stations: Badary (BD), Effelsberg (EF), Hartebeesthoek (HH), Irbene (IR), Jodrell Bank - Lovell (JB1), Jodrell Bank - Mark2 (JB2), Medicina (MC), Noto (NT), Onsala (O8), Shanghai - 25 m (SH), Tianma - 65 m (T6), Svetloe (SV), Toru\'n (TR), Yebes (YS), Westerbork (WB), Zelenchukskaya (ZC). e-MERLIN outstations: Cambridge, Darnhall, Defford, Knockin, and Pickmere.} 
\end{tiny}
}
\end{center}
\end{table*}

During August 2021, \rs{} underwent a new eruption. It was the first nova outburst ever detected at very high energies (E>100\,GeV) mapped by Cherenkov telescopes \citep{hess2022,Acciari2022}. Prompt observations as well as monitoring campaigns were triggered throughout electromagnetic spectrum, including radio \citep{Sokolovsky2021, Williams2021, deRuiter2023, Nayana2024}, infrared \citep{Woodward2021}, optical \citep{Munari2021,Munari2022a,Azzolini2023,Molaro2023,Tomov2023}, X-ray \citep{Shidatsu2021, Page2022,Orio2023,Ness2023,Islam2024}, and $\gamma$-rays \citep{Cheung2022}. The interpretation of the data thus gathered benefited from the fact that similar instrumentation and observing strategies as those used for the 2006 eruption were also adopted in 2021, especially in monitoring the evolution in X-rays \citep[e.g.,][]{Osborne2011,Page2022} and imaging in the radio the expansion of the ejecta \citep[e.g.,][]{Obrien2006,Rupen2008,Munari2022,Giroletti2023}. Of notable benefit has been the extremely accurate astrometric position and proper motions of \rs{} provided by \textit{Gaia} DR3, such as in the case of the recent work on the nova V407 Cygni by \citet{Giroletti2020}.

Within this context, we have conducted a multifrequency and five-epoch very long baseline interferometry (VLBI) monitoring campaign with the European VLBI network (EVN) at 1.6 and 5\,GHz that covered \rs{} from about two weeks (late August) to up to two months (early October) after the 2021 explosion.  The 5\,GHz data for the central epoch (day +34) have already been presented and modeled in 3D by \citet{Munari2022} in conjunction with results from high-resolution optical spectroscopy, while a preliminary overview on the whole set of EVN observations has been provided by \citet{Giroletti2023}.  In this paper, we publish the complete set of results from the full EVN monitoring that allows us to characterize, with an unprecedented level of detail, the evolving source structure and the physical conditions of the surrounding medium in which the nova ejecta propagated.

Assuming the \textit{Gaia} distance of 2.7 kpc, 1\,mas corresponds to 2.7 AU.  Throughout the paper the radio spectral index $\alpha$ is defined as $S_\nu \propto \nu^{\alpha}$, with $S_\nu$ representing the flux density at frequency $\nu$, and all angles are measured from north to east. Finally, we used the start of the eruption as the reference epoch, $T_0$, which has been derived to be 2021 August 08.50 ($\pm$0.01) UT by \citet{Munari2021}.


\section{Observations and data calibration} \label{sect_observations}

We monitored \rs{} from 2021 August 22 ($T_0$ + 14 days) to 2021 October 12 ($T_0$ + 65 days) with the EVN and e-MERLIN at 1.6 and 5\,GHz. The observations were organized into ten observing sessions grouped into five epochs, with each epoch consisting of two 8-hour runs on consecutive days at 5 and 1.6\,GHz, respectively (see Table\ref{table_obs}).

The EVN participating stations were: Badary, Effelsberg, Hartebeesthoek, Irbene, Jodrell Bank, Medicina, Noto, Onsala, Shanghai, Svetloe, Torun, Yebes, Westerbork, Zelenchukskaya. The last six observing runs also feature the e-MERLIN outstations in Cambridge, Darnhall, Defford, Knockin, and Pickmere.  Observations were performed in dual polarization mode with a bandwidth of 128\,MHz (divided into $4 \times 32$ MHz-wide sub-bands) at 1.6\,GHz and 256\,MHz (divided into $8 \times 32$ MHz-wide sub-bands) at 5\,GHz.  We used a standard phase-reference observing scheme with 4 min on the target and 40 seconds on the calibrator J1745-0753.

The data correlation was performed in real time with the software FX-kind correlator (SFXC) at the Joint Institute for VLBI ERIC (JIVE, the Netherlands) \citep{Keimpema2015}.  The a priori data reduction was carried out with the JIVE EVN pipeline based on ParselTongue \citep{Kettenis2006} and AIPS \citep{Greisen2003}.  We used the \texttt{DIFMAP} software package \citep{Shepherd1997} to inspect and edit the calibrated data downloaded from the EVN Data Archive\footnote{\url{http://archive.jive.nl/scripts/portal.php}} as well as for the imaging and self-calibration procedures. In more detail, for the imaging process we used the classical inverse modeling CLEAN algorithm, implemented in the \texttt{DIFMAP}, that represents the source structure as a collection of point sources and whose location and flux densities are iteratively determined \citep{Hogbom1974}.
We note that the two observing runs of Epoch IV (2021 September 26 and 27) were strongly affected by a hardware problem with e-MERLIN and were excluded from the analysis presented in this paper. Additional details about the observations can be found in \citet{Munari2022} and \citet{Giroletti2023}.

\section{Results} \label{sect_results}

 \begin{figure*}
\sidecaption
  \includegraphics[width=6.4cm]{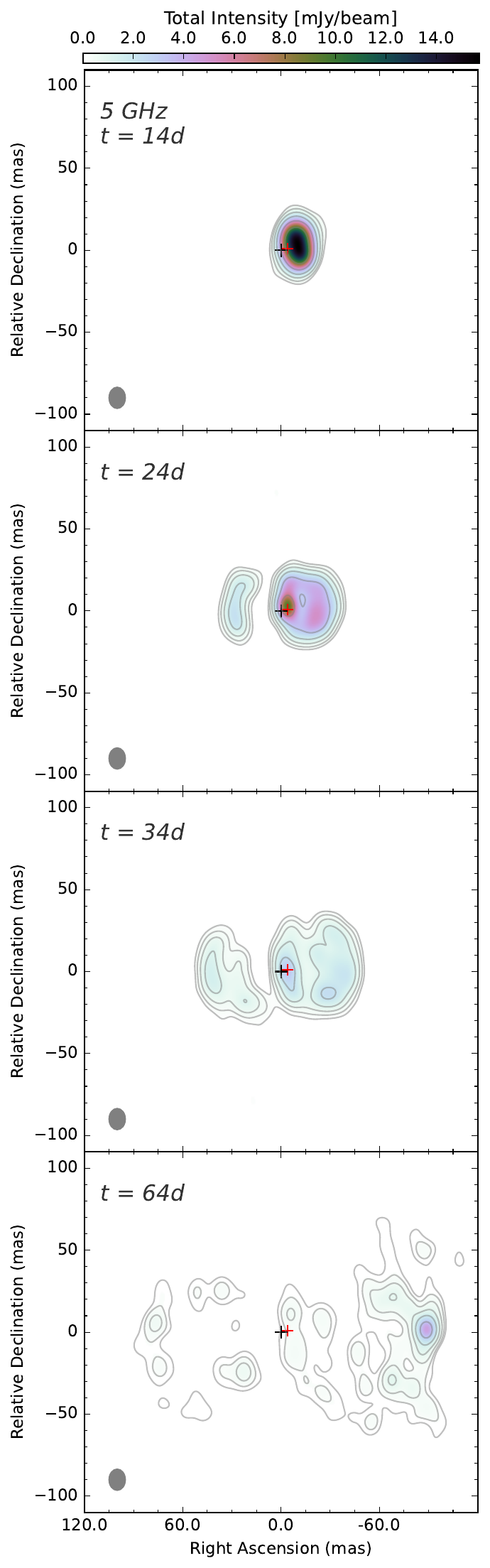} 
  \includegraphics[width=6.1cm]{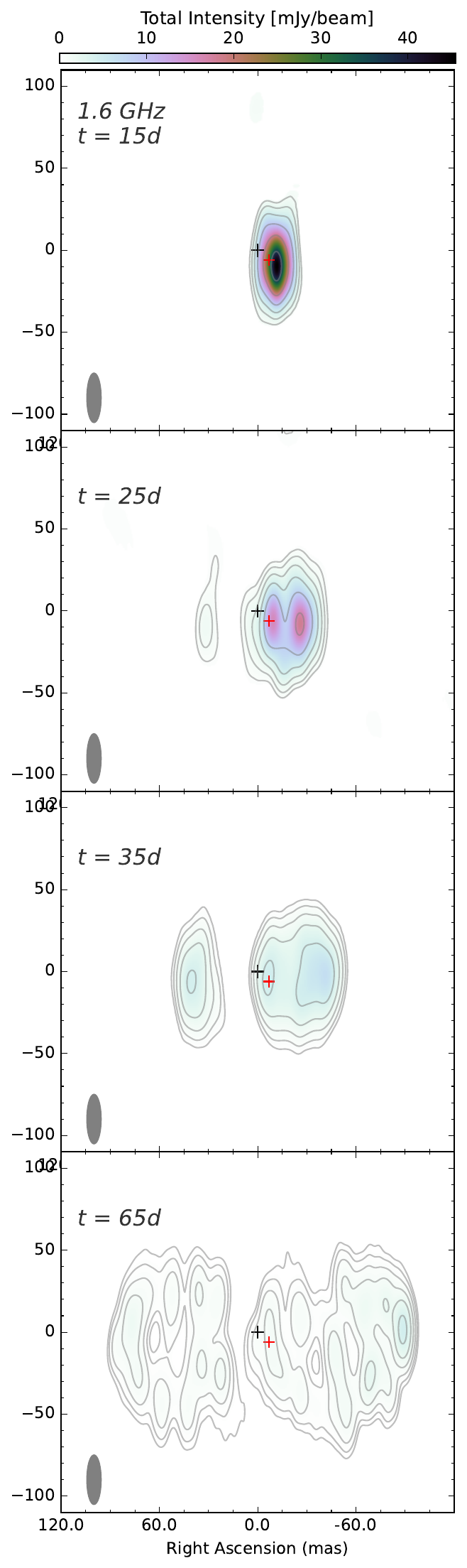} 
     \caption{Natural-weighted total intensity images at 5\,GHz (left frame) and 1.6\,GHz (right frame), centered at the \textit{Gaia} DR3 astrometric position for \rs{} (RA $17$h$50$m$13.1610$s, Dec $-06$d$42' 28.61026''$), corrected for proper motion (indicated by a black cross whose size represents the position uncertainty magnified by a factor of 40 for visualization).  The observing epoch, in units of days after $T_0$, is indicated in the top-left corner of each image. All 5\,GHz images (left panel) are displayed with a Gaussian taper of 0.5 at 10 M$\lambda$ and convolved with a beam of 9.8 mas $\times$ 12.8 mas at 0$^{\circ}$. All 1.6\,GHz images (right panel) are convolved with a beam of 8.5 mas $\times$ 30.0 mas at 0$^{\circ}$. This restoring beam is displayed in the bottom-left corner of each image. The color scale and the overlaid contours represent the total intensity emission, with the lowest contour at 2\% of the map peak (see Table \ref{table_obs}) and the following contours a factor of two higher. The red cross indicates the average core position (see Sect.~\ref{sect_exp_speed}).}
     \label{img_images}
\end{figure*}

\subsection{Images and source structure} \label{sect_images}

In the left panel of Fig.~\ref{img_images}, we show the 5\,GHz images obtained for the epochs I, II, III, and V (14, 24, 34, and 64 days after $T_0$, respectively). All images have been convolved with a beam of 9.8 mas $\times$ 12.8 mas at 0 deg, resulting from the average beam sizes across all the epochs, and are displayed with natural weights and a Gaussian taper of 0.5 at 10 M$\lambda$. The 1.6\,GHz images, obtained for the epochs I, II, III, and V (15, 25, 35, and 65 days after $T_0$, respectively), are reported in the right panel of Fig.~\ref{img_images}. All 1.6\,GHz images have been convolved with a beam of 8.5 mas $\times$ 30.0 mas at 0 deg, resulting from the average beam sizes across all the epochs, and are displayed with natural weights. All of the images are centered at the \textit{Gaia} DR3 astrometric position for \rs{}, corrected for proper motion (highlighted with a black cross at the center of each image).

\begin{figure}
\begin{center}
\includegraphics[bb = 8 9 575 455, width= 0.41\textwidth, clip]{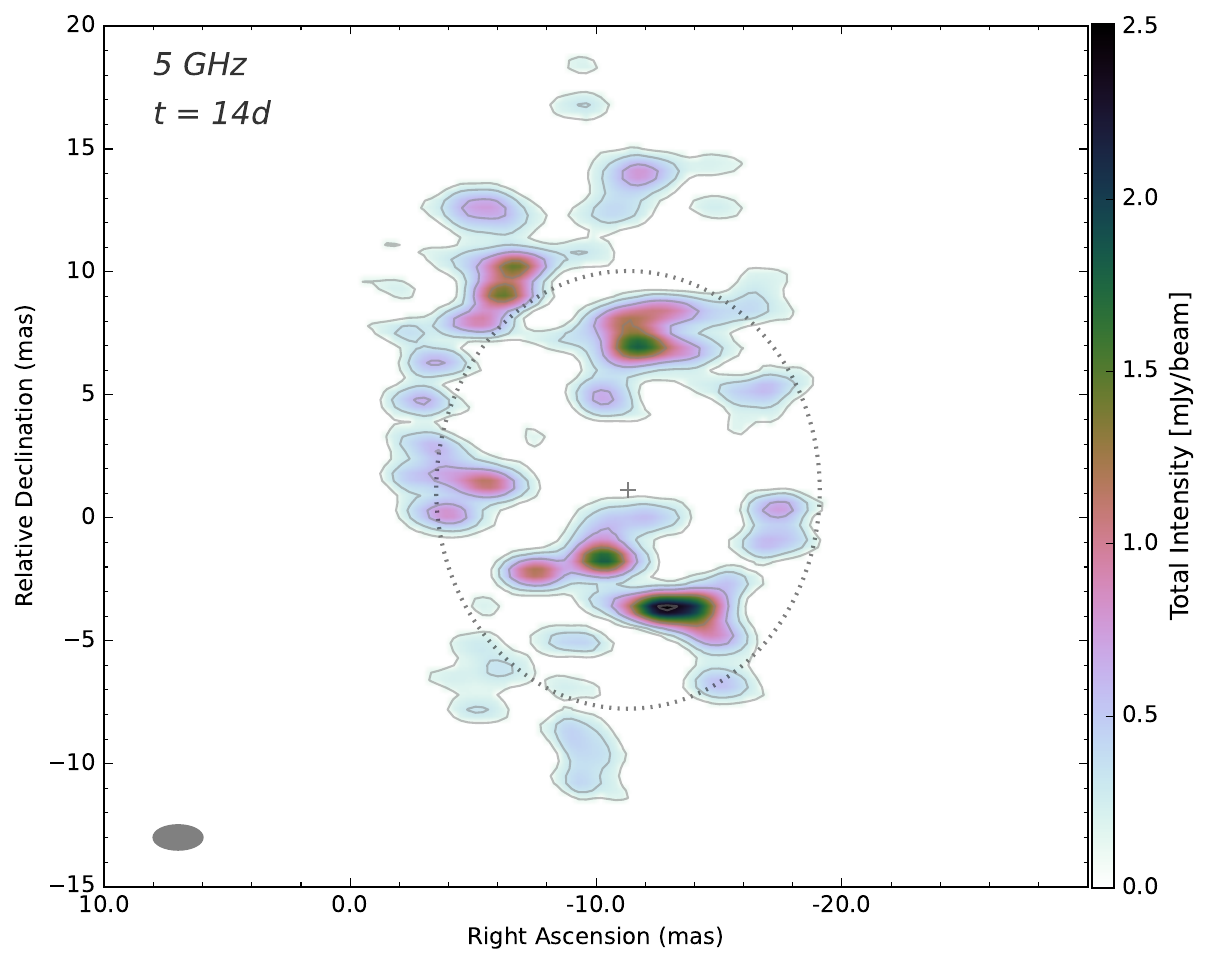}  
\end{center}
\caption{Full-resolution 5\,GHz total intensity image during epoch I (14
days after $T_0$) with relative coordinates referring to the \textit{Gaia} DR3 astrometric position  of \rs{}, corrected for proper motion. The convolving beam, displayed in the bottom-left corner of the image, is 1.0 mas $\times$ 2.0 mas at $90^{\circ}$. The color scale and the overlaid contours represent the total
intensity emission, with the lowest contour at 5\% of the map peak (see Table \ref{table_obs}) and the following contours a factor of two higher.  The gray dots represent the best-fitting ellipse obtained by \citet{Obrien2008} at the same frequency and epoch after the 2006 outburst.}
\label{img_full_res}
\end{figure}

The images at both frequencies reveal an evolving source morphology expanding in the east-west direction, that during the last observing epoch (64-65 days after $T_0$) reaches a total extension of $\sim200$ mas ($\sim540$ AU).  As it is apparent since epoch II (24-25 days after $T_0$), at both frequencies, the images reveal a triple-component source structure, with a compact central component (core) and two elongated lobe structures propagating at opposite sides of the core. We note that the compact core is located a few milliarcseconds from the \textit{Gaia} DR3 position (see Sect.~\ref{sect_exp_speed} for more details). The flux density depression between the eastern lobe and the rest of the source is interpreted by \citet{Munari2022} as being due to absorption from ionized gas in the so-called density enhancement on the orbital plane (DEOP).

The western lobe is brighter than the eastern one and has a circular shape that is consistent with what was found in the images at similar observing frequencies from the previous 2006 nova event \citep{Obrien2006, Rupen2008}. On the other hand, the arc-like structure observed on the eastern side is assumed to represent the outer edge of the eastern lobe that emerges from behind the DEOP. 
Within this scenario, we interpret the compact structure that can be observed during the first epoch (14-15 days from $T_0$), at both frequencies, as the western lobe, with the core in the background and the eastern lobe behind the DEOP.

\subsection{First epoch 5\,GHz full-resolution image} \label{sect_full_res}

To better investigate the structure at the early stages of evolution, we show in Fig.~\ref{img_full_res} the total intensity image of epoch I (14 days after $T_0$) at 5\,GHz in full resolution centered at the \textit{Gaia} position. The image was obtained by adopting uniform weights and a convolving beam of 1.0 mas $\times$ 2.0 mas at $90^{\circ}$.  The radio emission, which could hardly be resolved in the image obtained with natural weights, is now found to be distributed in a complex and elongated morphology. The total extension is over $\sim30$ mas in the north-south direction and $\sim20$ mas in the east-west direction. Several spots of enhanced brightness are distributed mostly along the outer rim of this roughly elliptical region. These spots may mark locations of harder collision between the fast ejecta and the preexisting slow circumstellar material, a situation reminiscent of what \citet{Chomiuk2014} first observed in Nova Mon 2012.

At the same observing frequency, a similar elongated source structure was reported by \citet{Obrien2008} with the Very Long Baseline Array (VLBA) 14 days after the 2006 outburst. For comparison, in Fig.~\ref{img_full_res}, we report the best-fit ellipse (gray dots) obtained by \citet{Obrien2008}, which has a semi-major axis of $8.9$ mas and an axial ratio of $1.14$ mas.

\subsection{Lobe expansion speed} \label{sect_exp_speed}

To represent and quantify the emission from the central core that emerges from the second epoch on, we modeled the brightness distribution with a 2D circular Gaussian component in the visibility plane with the
\texttt{modelfit} procedure in \texttt{DIFMAP}. We defined a reference core position by averaging the core Gaussian component position during epochs II and III, when the core region is compact, bright, and clearly recognizable at both frequencies. The average core position is indicated in all images in Fig.~\ref{img_images} with a red cross at a distance from the \textit{Gaia} DR3 position of $\sim4$ and $\sim10$\,mas at 5 and 1.6\,GHz, respectively. The size of the expanding lobes was determined as the distance of the outermost lowest total intensity contour from the average core position (see Table~\ref{table_lobe_size_speed}). We note that the orbital separation of \rs{} amounts to $\sim1.5$~AU or $0.6$ mas at the \textit{Gaia} distance.

By considering the extension of the two lobes from the core region during the different epochs (Fig.~\ref{img_exp_speed}) and assuming zero separation at $T_0$, we obtained their projected expansion speeds by means of a least-squares regression (column 6 in Table
\ref{table_lobe_size_speed}). We note that for the uncertainties on the distance from the core, we assumed a conservative value equal to the beam minor axis. Since during epoch I (14-15 days after $T_0$) the eastern lobe is not yet visible and we cannot easily disentangle the western lobe from the core region, we did not consider this epoch when estimating the lobe expansion rate. 

For the western lobe, we obtained an average projected expansion speed of $6840 \pm 370$ km s$^{-1}$ at 5\,GHz and $6660 \pm 100$ km s$^{-1}$ at 1.6\,GHz. For the eastern lobe, we obtained an average projected expansion speed of $6830 \pm 340$ km s$^{-1}$ at 5\,GHz and $7100 \pm 540$ km s$^{-1}$ at 1.6\,GHz. The projected expansion velocity of the two lobes is therefore the same well within their combined errors.

\subsection{Light curves and spectral index}

To quantify the emission from the different source components, we used the following approach. We determined the source total flux density by summing up the flux density over all the clean components in \texttt{DIFMAP}, while the core flux density was provided by the 2D circular Gaussian component that represents the core region (see Sect.\ref{sect_images}). The eastern lobe can be clearly isolated from the overall source structure, and its flux density was obtained by summing up the flux density over all the clean components in this region. Lastly, the western lobe flux density was determined by subtracting the emission of the eastern lobe and the core from the total flux density.

\begin{figure*}
\begin{center}
\includegraphics[scale=0.45]{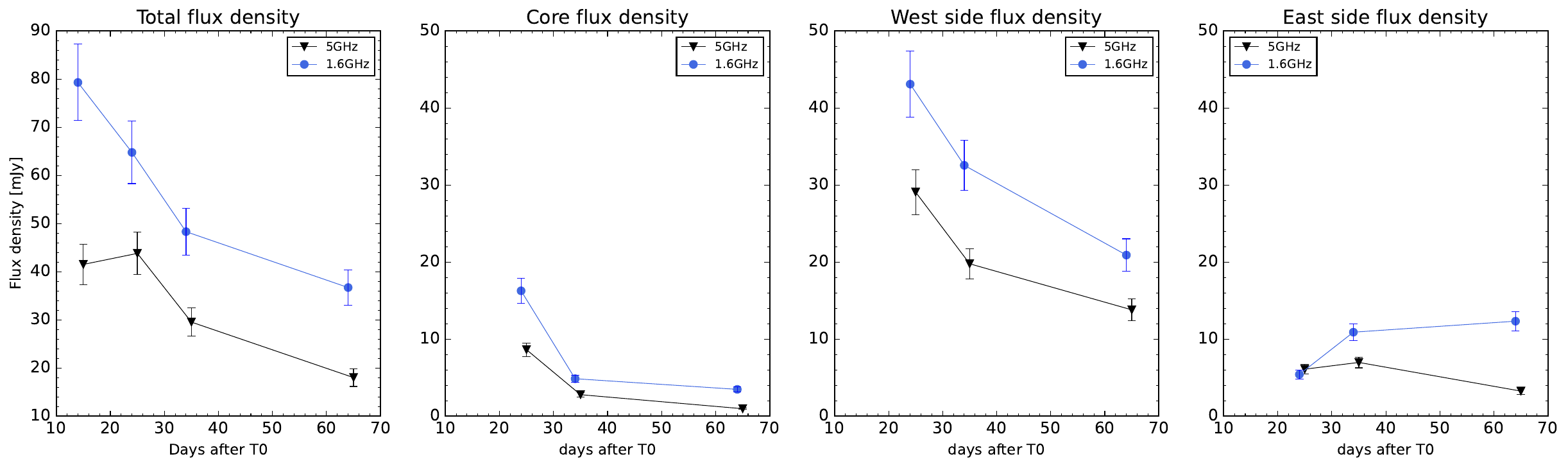}  
\end{center}
\caption{Light curves at 1.6\,GHz (blue circles) and 5\,GHz (black
triangles) for the whole source (first frame), the core (second frame), the western lobe (third frame), and the eastern lobe (fourth frame). All values are reported in Table \ref{table_flux_density}.}
\label{img_lc}
\end{figure*}

In Fig.~\ref{img_lc}, we report the resulting light curves. The flux density values are reported in table~\ref{table_flux_density}, with the uncertainties calculated by considering a calibration error of 10\% of the flux density and a statistical error equal to the map rms noise multiplied by the number of beams in the area over which the flux density was determined.

The total source emission (first frame in Fig.~\ref{img_lc}) shows an overall decreasing trend with time, with a $\sim50\%$ drop from epoch I ($\sim80$ mJy and $\sim40$ mJy at 1.6\,GHz and 5\,GHz, respectively) to Epoch V ($\sim40$ mJy and $\sim20$ mJy at 1.6\,GHz and 5\,GHz, respectively).
A similar decreasing trend in the flux density with time was also found at both frequencies for the core and the western lobe (second and third frame in Fig.~\ref{img_lc}, respectively).  We also note that the western lobe is brighter than both the core and the eastern lobe.
In contrast, for the eastern lobe, the flux density tends to increase with time, as the outer part of the lobe itself emerges behind the DEOP (fourth frame in Fig.~\ref{img_lc}). During the last observing epoch at 1.6\,GHz, the eastern lobe flux density decreases with respect to the previous two epochs, as the source becomes more diffuse.

\begin{figure}
\begin{center}
\includegraphics[bb = 8 8 336 327, width= 0.4\textwidth, clip]{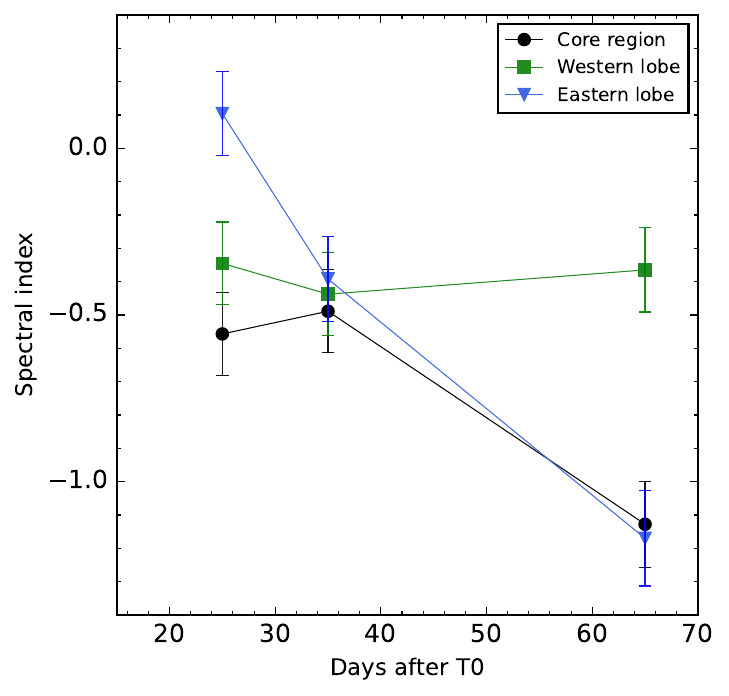}
\end{center}
\caption{Spectral index at 1.6 - 5\,GHz for the core region (black circles), western lobe (green squares), and eastern lobe (blue triangles). All values are reported in Table \ref{table_spectral_index}.}
\label{img_si}
\end{figure}

The spectral index for the different source components is reported in Fig.~\ref{img_si} and Table \ref{table_spectral_index}. The western lobe (green squares) has a slightly negative spectral index, with an average value of $-0.38\pm 0.12$ across the entire observing period, covering about two months after $T_0$. The core region (black circles) has an average value of $-0.50\pm 0.12$ until epoch III (34-35 days after $T_0$), and it then becomes steeper ($-1.13\pm 0.13$) during the last epoch (64-65 days from $T_0$). On the other hand, the eastern lobe during epoch II (24-25 days after $T_0$) shows a slightly positive spectral index (blue triangles in Fig.~\ref{img_si}), with a value of $0.10\pm 0.13$. In the following epochs, the eastern lobe spectral index follows a trend similar to that of the western lobe: It becomes slightly negative during epoch III (34-35 days after $T_0$), to a value of $-0.39\pm 0.13$, and later steepens to a value of $-1.17\pm 0.14$ during the last epoch (64-65 days from $T_0$). The uncertainties on the spectral index were calculated from the error propagation theory (see \citealt{Lico2012}).

\subsection{Core region brightness temperature}

For the core region, which we represent by means of 2D circular Gaussian components across the different epochs (see Sect.~\ref{sect_exp_speed}), we estimated the observed brightness temperature ($T_{B}^{obs}$) by using the following equation \citep[e.g.,~][]{Tingay1998}:
\begin{equation}
T_{B}^{obs}=1.22 \times 10^{12} \frac{S_{core}}{\theta_{maj} \theta_{min}\nu^2},
\end{equation}
\noindent
with $S_{core}$ being the fitted core flux density in Jy, $\theta_{maj}$ and $\theta_{min}$ respectively as the full width at half maximum (FWHM) of the major and minor axes of the 2D Gaussian component in milliarcseconds, and $\nu$ as the observing frequency in gigahertz. The FWHMs of the Gaussian components during epochs II, III, and V are respectively $1.6$, $4.2$, and $12.7$\,mas at 5\,GHz and $4.4$, $5.9$, and $15.5$\,mas at 1.6\,GHz.

During epoch II (24-25 days after $T_0$), III (34-55 days after $T_0$), and V (64-65 days after $T_0$), we found a brightness temperature on the order of $7.0 \times 10^7$\,K, $7.8 \times 10^6$\,K, and $3.0 \times 10^5$\,K at 5\,GHz and $4.0 \times 10^8$\,K, $6.7 \times 10^7$\,K, and $4.6 \times 10^6$ at 1.6\,GHz, respectively. This clearly supports a nonthermal origin of the radio emission, consistent with spectral index $<-0.1$.

\begin{figure}
\begin{center}
\includegraphics[bb = 8 8 346 333, width= 0.4\textwidth, clip]{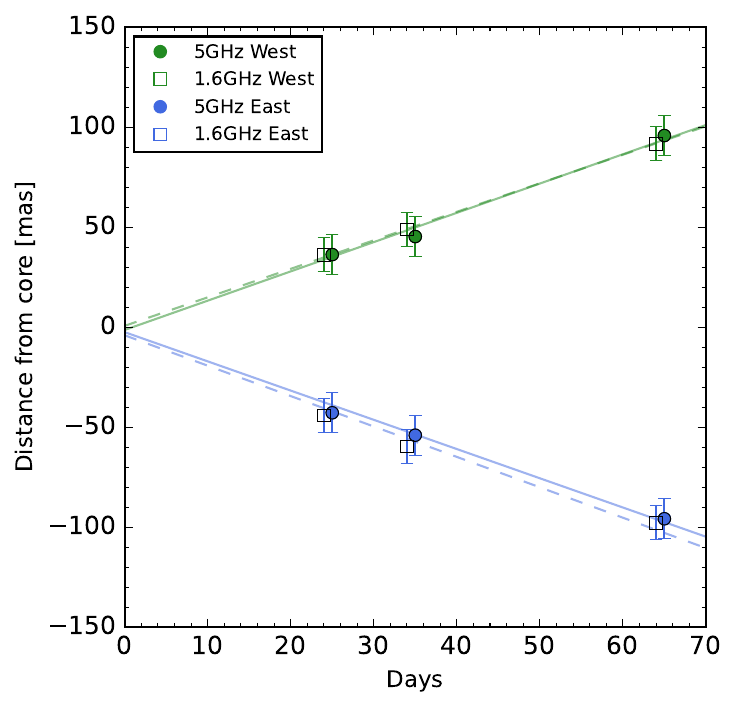}  
\end{center}
\caption{Angular separation evolution with time of the
eastern (blue) and the western (green) lobes at 1.6\,GHz (empty squares) and 5\,GHz (filled circles). The solid and dashed lines represent the best least-square regression fit for the 5 and 1.6\,GHz data, respectively. }
\label{img_exp_speed}
\end{figure}

\begin{table}
\begin{center}
\begin{tiny}
\caption{Lobe extension during the different observing epochs (columns 3 - 5) and expansion speed (column 6).}
\label{table_lobe_size_speed}   
\setlength{\tabcolsep}{3.2pt}
\renewcommand{\arraystretch}{1.0}     
\begin{tabular}{ccrrrc}  
\hline
\noalign{\smallskip}

 & Lobe & Epoch II & Epoch III & Epoch V & Projected \\
  &      &          &           &         & expansion speed\\
  & & mas (AU) & mas (AU) & mas (AU) & km s$^{-1}$\\
\noalign{\smallskip}
\hline    
\noalign{\smallskip}
5\,GHz & West & 36.3  (98) & 45.3 (122) & 95.8 (259) & 6840 $\pm$ 370 \\
  & East & 42.8 (116) & 54.0 (146) & 95.8 (259) & 6830 $\pm$ 340 \\
1.6\,GHz & West & 36.1  (97) & 48.8 (132) & 91.6 (247) & 6660 $\pm$ 100\\
    & East & 44.1 (119) & 59.7 (161) & 97.9 (264) & 7100 $\pm$ 540\\   
\hline                                  
\end{tabular}
\end{tiny}
\end{center}
\end{table}

\section{Discussion} \label{sect_discussion}

With each new eruption, understanding of the outburst mechanism of \rs{} steadily improves due to ever more accurate data and comparison to previous events. Thanks to the radio imaging of the 2021 eruption presented in this paper, the evolving morphology of the expanding bipolar ejecta can be characterized in great detail, and we can determine the physical conditions of the surrounding medium in which they propagate.

\subsection{The evolving source morphology} 
\label{sect_discussio_evolving_morph}

From the high-resolution radio observations presented in this work, we found an expanding source structure consisting of three main components: a central and compact core region located within a few milliarcseconds from the \textit{Gaia} DR3 position \citep{Gaia2016, Gaia2023} and two lobes expanding in the east-west direction on opposite sides of the central binary. This evolving source structure is notably consistent with the images at similar frequencies for the past \rs\ nova event in 2006 reported by \citet{Obrien2006, Rupen2008, Sokoloski2008}. However, thanks to the accurate astrometric position provided by the \textit{Gaia} DR3 position, a key improvement from the 2021 data is the realization that the ring-like structure at early times is actually the western lobe, as the binary system is located just to the east of it.
This provides firm support for the bipolar outflow expansion scenario. Furthermore, as we argue in \citet{Munari2022}, the images obtained from this observing campaign clearly reveal that the gravitational pull of the WD channels most of the mass lost by the RG toward the orbital plane and forms the so-called DEOP. The confinement of the ejected material into the two observed bipolar outflow is produced by the combined effect of the DEOP and the accretion disk around the WD \citep[e.g.,][]{Orlando2012}. Within this scenario, the western side
represents the approaching lobe that is located in the foreground of the DEOP, while the eastern side represents the receding lobe expanding behind the DEOP. Considering that the nuclear burning at the surface of the WD continued uninterrupted to day +82 (i.e., well past the latest radio observation \citep{Munari2022a,Page2022}), we assumed full ionization within DEOP over its radial extent explored by our radio observations.

\begin{table}
\begin{center}
\begin{tiny}
\caption{Flux densities of the different source regions.}
\label{table_flux_density}   
\setlength{\tabcolsep}{3.2pt}
\renewcommand{\arraystretch}{1.0}     
\begin{tabular}{llcccc}  
\hline
\noalign{\smallskip}
&& Epoch I & Epoch II & Epoch III & Epoch V \\
&& mJy & mJy & mJy & mJy \\
\noalign{\smallskip}
\hline    
\noalign{\smallskip}
5\,GHz & Total & 41.5 $\pm$ 4.2 & 43.8 $\pm$ 4.4 & 29.5 $\pm$ 3.0 & 18.0 $\pm$ 1.8 \\
& Core & - &  8.6 $\pm$ 0.9 & 2.8 $\pm$ 0.3 & 1.0 $\pm$ 0.1 \\
& East & - &  6.1 $\pm$ 0.6 & 7.0 $\pm$ 0.7 & 3.2 $\pm$ 0.4 \\
& West & - &  29.1 $\pm$ 2.9 & 19.8 $\pm$ 2.0 & 13.8 $\pm$ 1.4 \\
1.6\,GHz & Total & 79.3 $\pm$ 7.9 & 64.8 $\pm$ 6.5 & 48.3 $\pm$ 4.8 & 36.7 $\pm$ 3.7 \\
& Core & - &  16.3 $\pm$ 1.6 & 4.9 $\pm$ 0.5 & 3.5 $\pm$ 0.4 \\
& East & - &  5.4 $\pm$ 0.6 & 10.9 $\pm$ 1.1 & 12.3 $\pm$ 1.3 \\
& West & - &  43.1 $\pm$ 4.3 & 32.6 $\pm$ 3.3 & 20.9 $\pm$ 2.1 \\
\hline                                  
\end{tabular}
\end{tiny}
\end{center}
\end{table}

In more detail, this is the evolution process we can infer from the images presented in Fig.\ref{img_images}.  During the first epoch (14-15 days after $T_0$), at both 1.6 and 5\,GHz, we only observed the approaching western lobe, projected on the plane of the sky, with emission from the
central region absorbed by the dense and ionized gas and the western lobe laying in front.  Similarly, the eastern lobe is completely hidden from view by absorption from the densest and ionized inner regions of DEOP. As the western lobe further propagates with a circular shape, during epoch II (24-25 days from $T_0$) the central core emerges with a bright and compact structure and a brightness temperature on the order of $7 \times 10^8$\,K at 5\,GHz and $4 \times 10^8$\,K at 1.6\,GHz, indicating the nonthermal nature of the emission. A nonthermal origin for the whole radio emission on these scales\footnote{The analysis of the source structure evolution on larger scales by means of multifrequency e-MERLIN and Jansky Very Large Array observations will be presented in a separate work (Williams at al.~in prep.).} is also indicated by the spectral properties. The integrated spectral index in the core region and the western lobe are negative, with average values of $-0.50\pm 0.12$ and $-0.40\pm 0.12$, respectively, and the eastern lobe has a slightly positive spectral index ($0.10 \pm 0.13$), which is explained by the fact that as the eastern lobe leading edge emerges behind the DEOP, it is clearly visible at 5\,GHz, while at 1.6\,GHz the free-free absorption effect is still significant (the optical depth being $\tau_\nu \propto \nu^{-2.1}$) and its emission is still opaque.
We note that the source extension and morphology during epoch II (24-25 days from $T_0$) are in good agreement with the results for the 2006 outburst reported by \citet{Rupen2008} (20.8 and 26.8 days after the outburst) and \citet{Obrien2006} (21.5 and 28.7 days after the outburst) and obtained with the VLBA at 1.7 and 5\,GHz.

We were able to track the expanding source structure during the following two epochs that reaches an overall east-west extension of $\sim200$ mas ($\sim540$ AU) in the last observing epoch (64-65 days after $T_0$).  As the source structure expands and becomes more diffuse, it also becomes fainter, showing an overall decreasing flux density trend with time, both for the core region
and the western lobe. On the other hand, the flux density of the eastern lobe shows an increasing trend as it gradually emerges behind the DEOP (see Fig.~\ref{img_lc}). We note that in this last epoch, the core flux density has dropped by $\sim80\%$, and the brightest source region is found in the western lobe edge, and it maintains a less steep spectral index of $-0.36\pm0.13$. On the other hand, both the core region and the eastern lobe become more transparent, with spectral indexes of $-1.13\pm 0.13$ and $-1.17\pm 0.12$, respectively.  We note that during the last observing epoch, the core brightness temperature decreases as well, down to values of $3.0 \times 10^5$\,K at 5\,GHz and $5.0 \times 10^6$\,K at 1.6\,GHz, indicating the deceleration of the ejecta close to the equatorial plane, and therefore the densest regions of DEOP, is almost completed. The 1.6\,GHz images from epochs III (35 days after $T_0$) and V (65 days after $T_0$) are consistent with the images reported by \citet{Sokoloski2008} obtained 34 to 51 days after the 2006 outburst.


As it emerges from the analysis of the angular separation with time of the two lobes (see Fig.~\ref{img_exp_speed}) during the $14 \leq (T-T_0) \leq 65$ day interval probed by our observations, the overall evolving source structure is consistent with a linear expansion and has no hints of significant deceleration of the eastern and western lobes expanding perpendicular to the orbital plane. This is an indication that the density gradient perpendicular to the orbital plane is very steep and that the deceleration of the lobes occurred during the first two weeks after the explosion, namely before our observations started. By considering that during the first 34 days after the explosion the lobes expand with an average de-projected expansion speed of $\sim7750$ km s$^{-1}$, as estimated by \citet{Munari2022}, we can infer that there is no significant amount of mass higher than 50~AU above the equatorial plane and probably confined much closer to the plane than this upper limit. We also note that a similar constant expansion of \rs{} bipolar ejecta was found by \citet{Montez2022} from X-ray observations performed up to more than five years after the 2006 outburst.

\subsection{Density profile and total mass of DEOP}

If we assume intrinsic similarity and symmetry at any of the observing epochs between the two expanding lobes, the flux ratio between the western lobe seen in the foreground and the eastern lobe seen through the absorbing DEOP allows the radial dependence of density along the DEOP to be estimated.
To this aim, we adopted the following expression for the free-free optical thickness of ionized gas \citep{Mezger1967, Condon2016}:
\begin{equation}
\tau_\nu\approx 3.28 \times 10^{-7} \,\, \left(\frac{T}{10^4}\right)^{-1.35} \nu^{-2.1} \,\, n_e^2 \,\, s,
\end{equation}
\noindent
with the temperature $T$ in Kelvin, the frequency $\nu$ in gigahertz, the electron density $n_e$ in cm$^{-3}$, and the geometrical thickness $s$ in pc and assuming the same number of electrons and ions $n_e n_i \approx n_e^2$.

Being interested in the order of magnitude of the derived density, we considered the DEOP as being approximated by a plane-parallel slab of total $s$ thickness constant through the radial extent with an electron density $n_e$ dependent only on the radial distance and otherwise constant in the height direction. In the previous section, we estimated that at a height of 50~AU above the equatorial plane, the density within the DEOP drops to negligible values. Assuming the drop to be exponential with height (i.e., $\rho$(z)$\propto$$\rho$(0)e$^{-z/z0}$), the accumulated column density is equivalent to the case of constant density over a height of $\leq$12~AU. Accordingly, we adopted $s$=20~AU in Eq.(2). Finally, for de-projection purposes, we adopted the orbital inclination of $i$=54$^\circ$ derived by \citet{Munari2022}.

From Fig.~\ref{img_images} and its comparison with equivalent mapping of the 2006 eruption by \citet{Rupen2008}, the earliest appearance of the extincted eastern lobe occurs at 5\,GHz on day +20 at an angular distance of approximately 25 mas from the position of \rs{}, corresponding to a de-projected distance of $\sim$114~AU. Taking a brightness ratio of 100 at this epoch for the western lobe with respect to the eastern one, the electron density results in n$_e$(114~AU)=2$\times$10$^6$~cm$^{-3}$. At the time of the latest radio epoch on day +65, the brightness ratio at 5\,GHz has reduced to four for the apex of the lobes located at $\sim$87\,mas from \rs{}, corresponding to a de-projected distance of $\sim$400~AU from central binary and resulting in an electron density of n$_e$(400~AU)=9$\times$10$^5$~cm$^{-3}$.
An estimate of the electron density at the DEOP inner regions can be obtained from the recombination time following the initial UV flash at the very start of the outburst, for which \citet{Munari2021a} derived $n_e$=1$\times$10$^7$~cm$^{-3}$.
  
\begin{figure}
\begin{center}
\includegraphics[bb = 7 7 345 266, scale=0.65, clip]{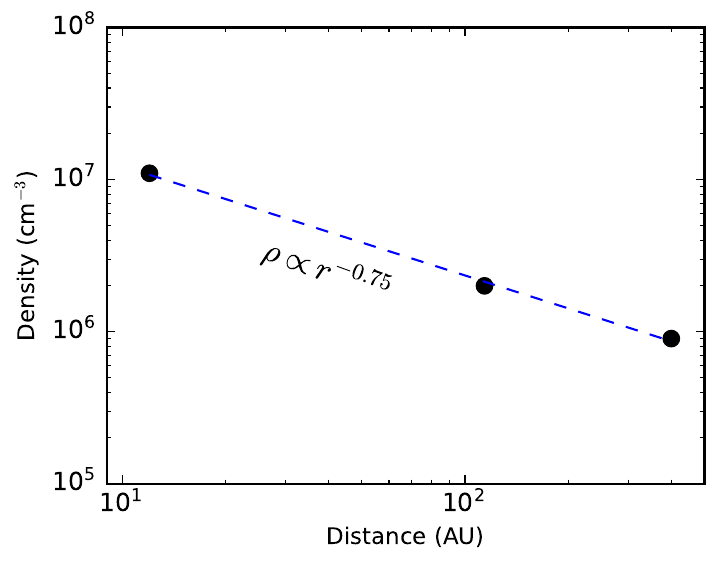}  
\end{center}
\caption{Dependence on radial distance from the central binary of the electron density within DEOP.}
\label{img_density}
\end{figure}

There is clear evidence of a radial dependence of the density going from 1.1$\times$10$^7$ ~cm$^{-3}$ close to the central binary to 2$\times$10$^6$ at 114~AU distance and to 9$\times$10$^5$~cm$^{-3}$ at 400~AU. The behavior is summarized in Fig.~\ref{img_density}, where the fitting curve represents the relation $\rho$(r) $\propto$ r$^{-0.75}$. 
Integrating the density along the DEOP volume between 10$\leq$r$\leq$175~AU leads to a total DEOP mass of
$\mathrm{M_{DEOP}} = 6.4 \times 10^{-6} ~~\mathrm{M_\odot}$.
Supposing the RG loses mass via a wind at a typical speed of 20 km s$^{-1}$ and that the vast majority of the mass lost ends up in the DEOP, the required RG mass-loss rate is
$\mathrm{\dot{M}_{\rm RG}=6.8 \times 10^{-8} ~~M_\odot/yr}$,
which compares well with standard estimates \citep[e.g.,][]{Livio1984} and argues in favor of the general soundness of the above reasoning.

To power an outburst every $\sim$20~yrs, the WD in \rs{} has to be massive and accrete at a high rate.  Theoretical models by \citet{Yaron2005} require accretion rates of $\dot{M}_{\rm WD}$$>$$10^{-8}$ M$_\odot$\,yr$^{-1}$ for a 1.4~M$_\odot$ WD and of $\dot{M}_{\rm WD}$$>$$10^{-7}$ M$_\odot$\,yr$^{-1}$ if the mass of the WD reduces to 1.25~M$_\odot$. These estimates are in good agreement with the results of X-ray observations of \rs{} in quiescence (performed during 2007 August 4 and 2008 February 26, respectively 537 and 744 days after the 2006 outburst) reported by \citet{Nelson2011} and that have placed an upper limit of $\dot{M}_{\rm WD}$$<$$2\times 10^{-8}$ M$_\odot$\,yr$^{-1}$ for \rs{}. Comparing with our estimate for the RG mass loss of $\mathrm{\dot{M}_{\rm RG}=6.8 \times 10^{-8} ~~M_\odot/yr}$, it is interesting to conclude that the WD in \rs{} is likely accreting around one-tenth of the mass lost by the RG
and that a great majority therefore goes to feed the circumstellar region and the DEOP in particular. This also has implications for the mass-reservoir constituted by the RG to keep powering new eruptions on the WD and, pending the mass-gain efficiency of eruption, the ultimate fate of the WD itself.


\subsection{High-resolution source morphology}

The full-resolution image at 5\,GHz for epoch I (14 days after $T_0$) reveals a
complex, granular-like source morphology (see Fig.~\ref{img_full_res}). Considering that during this epoch the total extension in the north-south direction is $\sim30$ mas, we obtained a transverse expansion speed of $\sim0.02$c.

A similar elongated source morphology was found with the VLBA at 5\,GHz 14 days after the 2006 outburst by \citet{Obrien2008}. The position, shape, and size of the brightness distribution best-fit ellipse reported in \citet{Obrien2008} are in good agreement with the total intensity source structure that we find 14 days after the 2021 outburst (gray dots in
Fig.~\ref{img_full_res}). This supports the idea that the way in which the RG wind refills the circumstellar space, which has been emptied by the expanding ejecta of the previous nova eruption, replicates in a self-similar manner at each new event, as the evolution at optical wavelengths of
each outburst is identical to all others.

\begin{table}
\begin{center}
\begin{tiny}
\caption{Spectral index values for the different source regions.}
\label{table_spectral_index}   
\setlength{\tabcolsep}{3.2pt}
\renewcommand{\arraystretch}{1.0}     
\begin{tabular}{cccc}  
\hline
\noalign{\smallskip}
Spectral index & Epoch II & Epoch III & Epoch V \\
\noalign{\smallskip}
\hline    
\noalign{\smallskip}
Core & $-0.56$ $\pm$ 0.12 & $-0.49$ $\pm$ 0.12 & $-1.13$ $\pm$ 0.13 \\
East & 0.10 $\pm$ 0.13 & $-0.39$ $\pm$ 0.13 & $-1.17$ $\pm$ 0.14 \\
West & $-0.35$ $\pm$ 0.12 & $-0.44$ $\pm$ 0.12 & $-0.36$ $\pm$ 0.13 \\
\noalign{\smallskip}
\hline    
\end{tabular}
\end{tiny}
\end{center}
\end{table}

A marked granular structure, although at a much closer distance to the WD, is supported by the X-ray observations at the onset of the supersoft phase during the 2006 outburst of \rs{} \citep{Osborne2011} and, to a lesser degree, also during the 2021 event \citep{Page2022}. Our EVN
observations suggest that such granularity is present also at scales much larger than implied by the X-ray observations.
It is beyond the goals of this paper to investigate the origin of such a clumpy structuring of the ejecta and of the circumstellar medium around \rs{}, but various mechanisms may be at work at different distances from the central binary involving inhomogeneities in the wind blowing off the WD during the burning phase as well as the turbulent nature of shocks propagating through the preexisting (and probably itself clumpy) slow circumstellar material.

\section{Conclusions} 
In this work, we have presented the results of a high-resolution VLBI radio monitoring of the recurrent and symbiotic nova \rs{} after the most recent outburst that occurred in August 2021. By means of multi-epoch EVN plus e-MERLIN observations at 1.6 and 5\,GHz from 14 to 65 days after the explosion, we characterized in great detail the evolving source morphology and the physical conditions of the surrounding medium in which the ejected material propagates. 

At both frequencies, we identified a triple-component source morphology with a compact central core region and two elongated bipolar outflow expanding linearly in the east-west direction with an average projected speed of $\sim7000$ km s$^{-1}$. 
We estimate a total mass of $6.4 \times 10^{-6} ~~\mathrm{M_\odot}$ for the DEOP, representing the density enhancement above the orbital plane that we reveal in our images as a flux density depression between the eastern lobe and the rest of the source. 
We found a clear radial dependence of the density within the DEOP that varies from 1$\times$10$^7$ ~cm$^{-3}$ in the near vicinity of the central binary to 9$\times$10$^5$~cm$^{-3}$ at $\sim400$~AU. 
Based on our results, we can firmly conclude that the majority of the mass lost by the RG companion ends up in the DEOP and the circumstellar region, with only about one-tenth of it being accreted by the WD.

\begin{acknowledgements}
We thank the anonymous referee for the insightful comments and suggestions that improved the quality of the manuscript. The European VLBI Network is a joint facility of independent European, African, Asian, and North American radio astronomy institutes. Scientific results from data presented in this publication are derived from the following EVN project code: RG012. The research leading to these results has received funding from the European Union's Horizon 2020 Research and Innovation Programme under grant agreement No. 101004719 (OPTICON RadioNet Pilot). e-MERLIN is a National Facility operated by the University of Manchester at Jodrell Bank Observatory on behalf of STFC.
This work has made use of data from the European Space Agency (ESA) mission {\it Gaia} (\url{https://www.cosmos.esa.int/gaia}), processed by the {\it Gaia}
Data Processing and Analysis Consortium (DPAC, \url{https://www.cosmos.esa.int/web/gaia/dpac/consortium}). Funding for the DPAC has been provided by national institutions, in particular the institutions participating in the {\it Gaia} Multilateral Agreement. RL, MG, UM acknowledge financial support from INAF 2022 fundamental research programme ob. fun. 1.05.12.05.16. BM acknowledges financial support from the State Agency for Research of the Spanish Ministry of Science and Innovation under grant PID2019-105510GB-C31/AEI/10.13039/501100011033 and through the Unit of Excellence Mar\'ia de Maeztu 2020--2023 award to the Institute of Cosmos Sciences (CEX2019- 000918-M).
\end{acknowledgements}

\end{document}